# The Streaming Limit of Solar Energetic-Particle Intensities

**Donald V. Reames[1], Chee K. Ng[2]**

[1]Institute for Physical Science and Technology, University of Maryland, College Park, MD 20742-2431 USA, email: dvreames@umd.edu

[2]George Mason University, Fairfax, VA 22030, USA: cng2@gmu.edu

**Abstract** As solar energetic particles (SEPs) stream outward along the interplanetary magnetic field after acceleration by shock waves near the Sun, their intensities are limited by scattering against self-generated Alfvén waves, trapping the particles near their source. This streaming limit varies with the magnetic rigidity of the particle and with distance from the source. Pitch-angle coupling can cause higher-energy protons to suppress the intensities of lower-energy ions causing flattened low-energy spectra on the early SEP intensity plateau. At sufficiently high energies, particle flow and wave trapping of particles weakens and the SEP spectra steepen, forming spectral "knees".

D. V. Reames and C. K. Ng

## 1. Introduction

Are there limits on the intensities of solar energetic particles (SEPs)? Some years ago, Reames (1990) noticed an apparent limit on the intensity of low-energy protons early in large SEP events, as seen in the left panel of Figure 1. Later in some of these events, the intensity increased by factors of 10 or 100 at the time of shock passage. At that time it was already known that particles streaming along a magnetic field amplify Alfvén waves (Stix 1962, 1992; Melrose 1980) and Alfvén-wave amplification on cosmic ray propagation and escape from the galaxy has been studied for many years (see review by Wentzel 1974). Furthermore, in theories of diffusive shock acceleration (*e.g.* Bell 1978; Lee 1983) these waves scatter subsequent particles, thus trapping them in the vicinity of the shock where they received further acceleration. Equilibrium is established between wave amplification and particle scattering, which reduces the streaming and hence the wave growth. Eventually, increasing the source of particles only increases the wave intensity and the scattering with no increase of the outflowing particle intensities seen by a distant observer. This is the streaming limit.

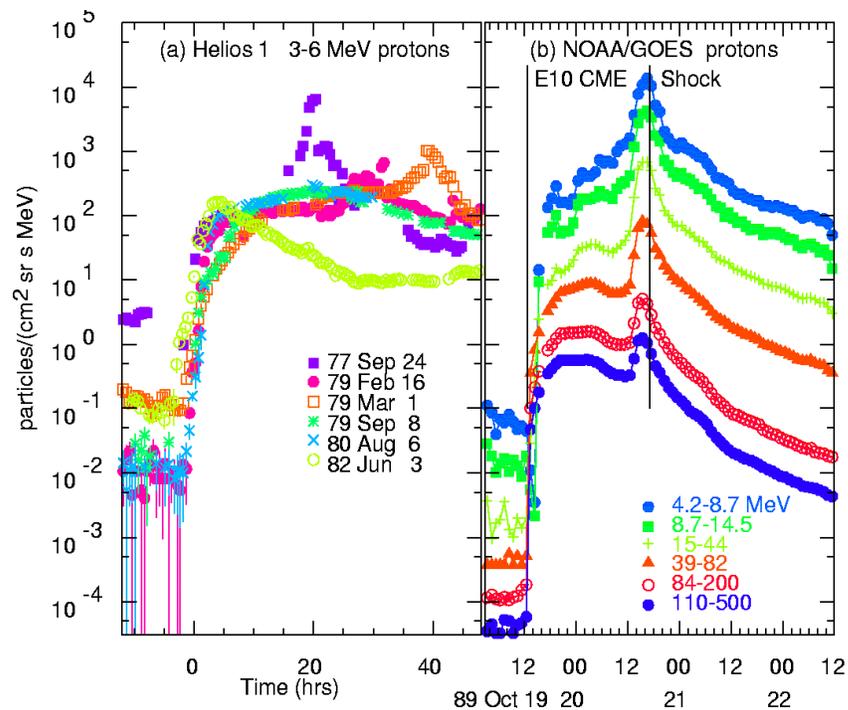

**Figure 1.** Panel (a) shows superposed intensity-time profiles of 3-6 MeV protons in several events with streaming-limited intensities early in the events (Reames 1990). Panel (b) shows similar limits as a function of energy in the large 1989 October 19 event. Intensities often peak at the time of shock passage at intensities that are 10-100 times the streaming limit.

If we are concerned with the radiation hazard of SEPs, it is higher-energy particles that will be of greatest interest; the right-hand panel of Figure 1 suggests limits for energy intervals up to 500 MeV. However, the streaming limit is a transport phenomenon; it does not apply *at* a shock peak. Furthermore, it is an equilibrium that may require time to establish, as we shall see.

## 2. Energy Dependence of the Streaming Limit

We attempted to define observational limits on the energy dependence of the streaming limit (Reames and Ng 1998) using data from an 11-year period from the *Geostationary Operational Environment Satellite* (GOES) of the *National Oceanic and Atmospheric Administration* (NOAA). Evidence of these limits is shown in Figure 2.

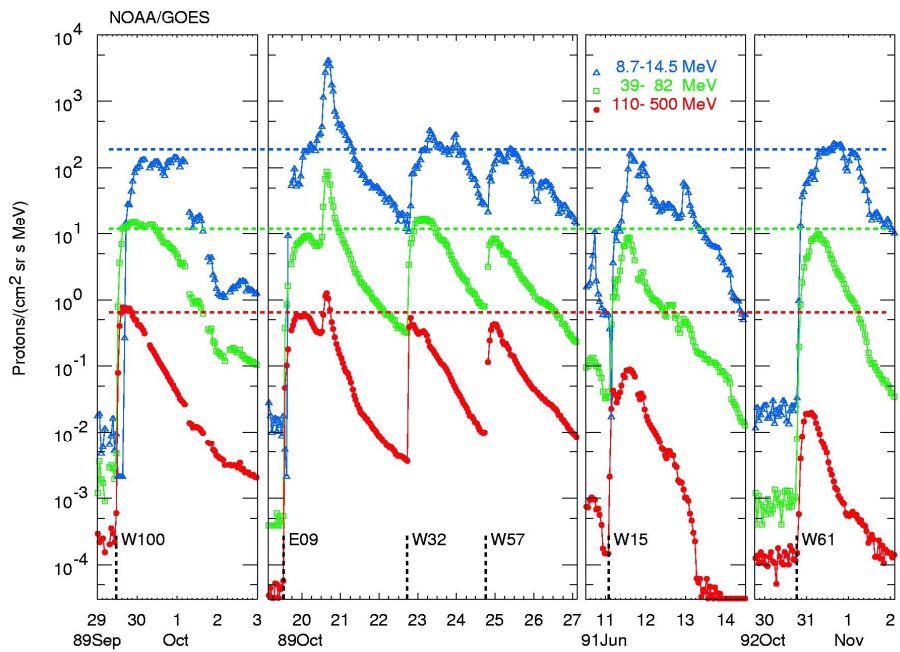

**Figure 2**. Intensity-time profiles of protons in three energy channels are shown for six large SEP events during solar cycle 22 as measured on the NOAA/GOES spacecraft. Streaming-limited intensity values for each energy channel are shown as dashed lines.

Figure 3 shows intensity distributions, *i.e.* the time spent at different logarithmically-spaced intensity levels, for three different proton energy intervals based upon over 11 years of GOES data. The intensity distribution for each energy interval has a power-law behavior up to the streaming limit. At that point they drop to a level where only intensities near shock peaks contribute.

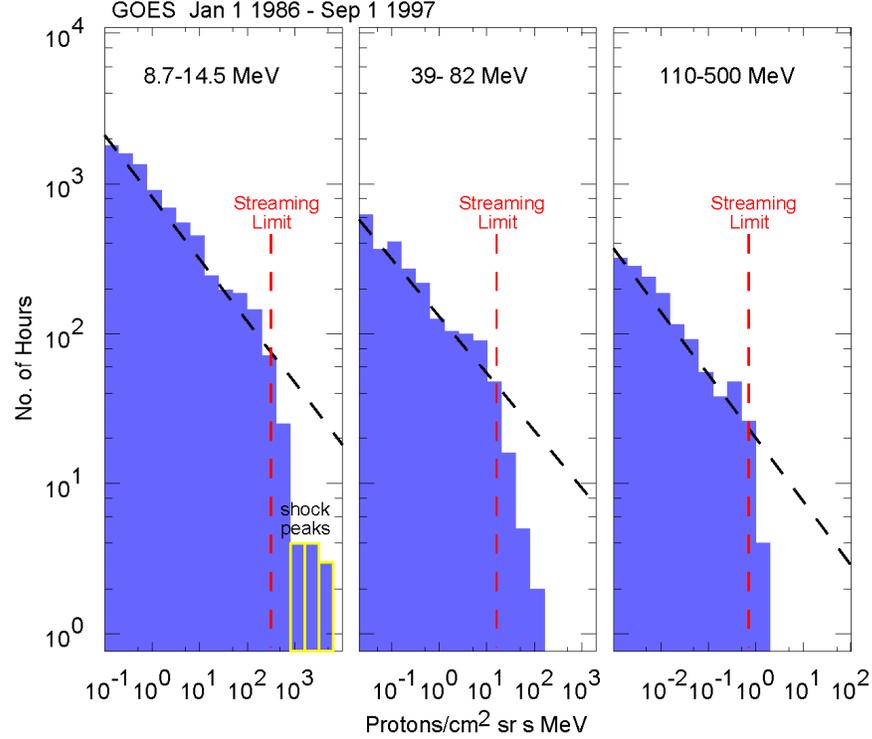

**Figure 3.** Over 11 years of GOES data in three energy intervals have been binned to show the number of hours spent an each intensity level. These intensity distributions show a power-law behavior up to the streaming limit. Intensity values above the streaming limit occur near shock peaks, which are not limited in intensity.

## 3. Wave Growth and the Streaming Limit

The amplification of Alfvén waves by streaming protons is discussed in textbooks on plasma physics (*e.g.* Melrose 1980; Stix 1992). Ions streaming along **B** resonate with Alfvén waves of wave number $k$:

$$k = \frac{B}{(\mu - V_A/v)P} \approx \frac{B}{\mu P} \quad (1)$$

where $V_A$ is the local Alfvén speed, $P=pc/Qe$ is the rigidity of a particle of charge $Q$, momentum $p$, and velocity $v$, and $\mu$ is the cosine of its pitch angle relative to **B**. The approximation is useful when $\mu \gg 0$ since we usually have $v \gg V_A$. The growth rate of the $\sigma$ mode of Alfvén waves is

$$\gamma_\sigma(k) = 2\pi^2 g_\sigma e^3 c V_A \iint d\mu dP \frac{P^3}{E^2} \frac{R^\sigma_{\mu\mu}}{(1-\mu V_\sigma/v)^2} \frac{\partial f_H}{\partial \mu} \quad (2)$$

where $g_\sigma = \pm 1$ for outward (inward) waves, $E$ is the total proton energy, $V_\sigma = W + g_\sigma V_A$, where W is the solar wind speed, $f_H$ is the proton phase-space

The Streaming Limit

density, and $R^{\sigma}_{\mu\mu}$ is the resonance function (see Ng and Reames 1995) that imposes the resonance condition (Equation 1) while allowing for resonance broadening. If we can ignore the slow motion of the waves relative to that of the particles, then the wave intensity of the $\sigma$ mode, $I_\sigma(k,r,t)$ obeys

$$\frac{\partial I_\sigma(k,r,t)}{\partial t} = \gamma_\sigma(k,r,t) I_\sigma(k,r,t) \qquad (3)$$

where we have explicitly shown the dependence upon space $r$ and time $t$, which may be extremely large. The pitch-angle diffusion coefficient for protons depends linearly upon the intensity of resonant waves.

Equilibrium shock acceleration theory (Bell 1978; Lee 1983, 2005) uses scattering on self-amplified waves to reflect particles back and forth across a shock as they gain an increment of velocity on each traversal. The spatial variation of proton intensities is shown in Figure 4 for increasing seed-particle injection at the shock for the Lee (1983) model which involved a planar shock. Local wave growth traps particles of increasing intensities near the shock while distant intensities are limited. Lee (1983) made the simplifying assumption that $\mu \approx 1$ so, at each energy, particles had their own unique resonant waves; a streaming limit was independently established at each energy.

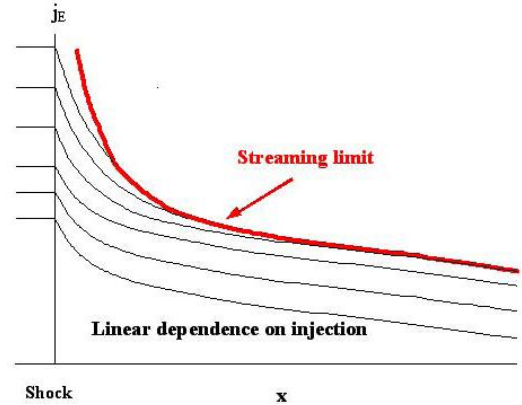

**Figure 4**. Proton intensities are shown for increasing seed-particle injection in the Lee (1983) shock model. At a distance, x, the intensity increases linearly at low injection, but reaches a limit at high injection because of increasing wave growth near the planar shock.

The time-dependent transport calculations of Ng and Reames (1994) derived the intensity of low-energy protons near 1 AU as a function of the intensity near the Sun as shown in Figure 5. Note that the 1-AU intensity actually peaks at the intensity value observed in the left panel of Figure 1. The streaming limit is an absolute value; it depends upon plasma parameters such as $V_A$, but there are no arbitrarily adjustable source parameters.

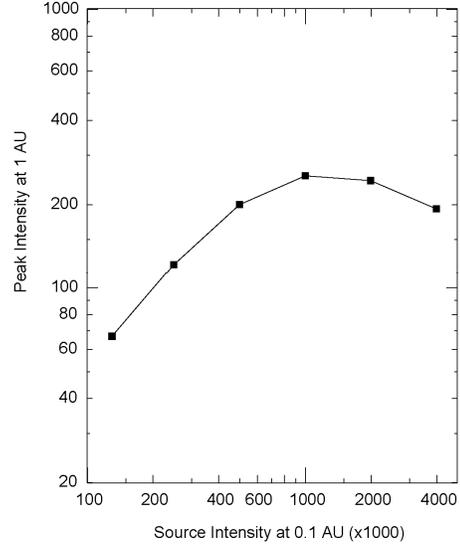

**Figure 5**. The intensity of 1 MeV protons near Earth is shown *vs.* the solar source strength at 0.1 AU for the transport calculations of Ng and Reames (1994). Note that at the proton streaming limit at 1 AU, the intensity is a factor of ~5000 or more higher at 0.1 AU.

## 4. The Plateau Spectrum and Pitch-Angle Coupling

When we examine the full energy spectra of ions in the early phase of large proton events, we often see spectra that are flattened at low energy as shown in Figure 6. The figure shows energy spectra of H and O from several of the largest SEP events in solar cycle 23. To understand the cause of the flattening we must consider the $\mu$ dependence in the resonance condition in Equation 1. As protons in the 10–100 MeV region stream out from the shock source they amplify resonate Alfvén waves, scatter, and begin to isotropize. Those at smaller $\mu$ amplify waves at higher $k$ which can scatter slower particles at lower $P$ (*e.g.* near 1 MeV) and $\mu \approx 1$ which are just beginning to arrive. Thus the low-energy ions are strongly suppressed by waves generated by higher-energy protons.

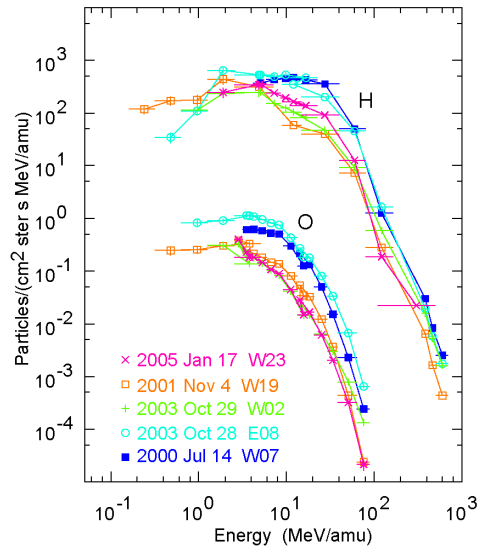

**Figure 6**. Energy spectra of H and O are shown for the early plateau region of five large SEP events (Reames and Ng, 2010).

The Streaming Limit

The coupling of different energies through the pitch-angle dependence is demonstrated by comparing the H spectra of the two large SEP events, both ground-level events (GLEs), shown in Figure 7. SEP events with too few high-energy protons are unable to generate adequate wave intensity to suppress the lower-energy protons.

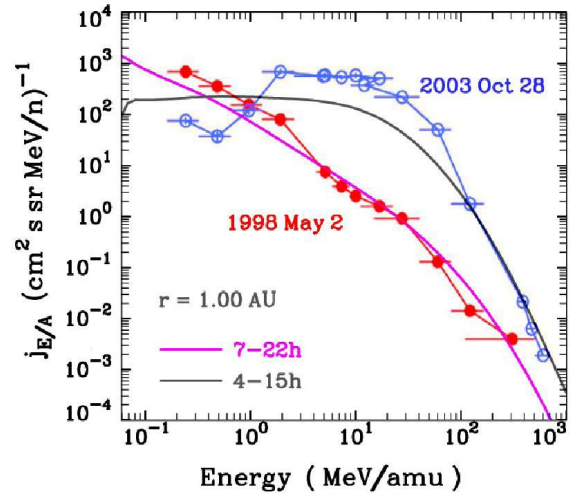

**Figure 7**. The proton energy spectra on the plateau of two large GLEs are compared. The October event has high intensities of 10-100 MeV protons that generate Alfvén waves that suppress the slower ~1 MeV protons as they emerge. The May event has a factor of ~100 fewer 10-100 MeV protons so the ~1 MeV protons are not suppressed. Numerical calculations of Ng, Reames, and Tylka (2012) confirm this behavior.

The rate of rise of the proton intensity can also be a factor in the establishment of equilibrium of the streaming limit as shown in Figure 8. The fast rise of high-energy protons in the SEP event of January 20, 2005 allows the intensity to exceed the equilibrium limit until there has been enough wave growth to establish the equilibrium. Events with slower evolution do not overshoot the streaming limit.

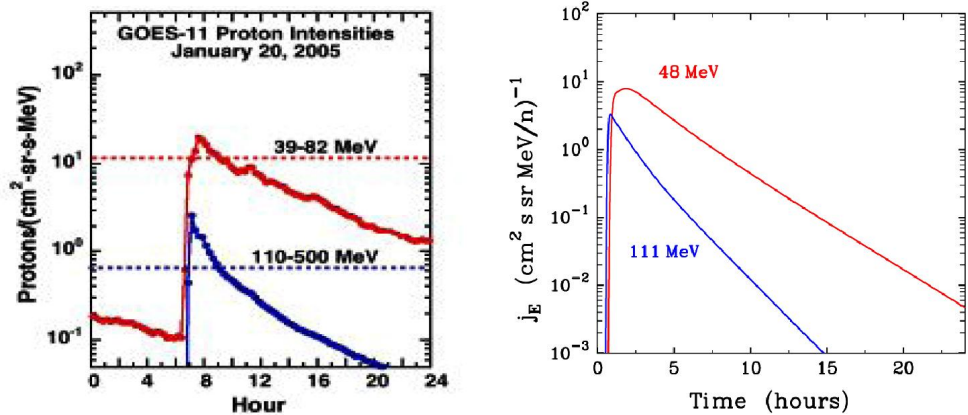

**Figure 8.** The left panel shows that intensities in the event of January 20, 2005 briefly exceed the expected streaming limits from Figure 2 (Mewaldt *et al.* 2007). The right panel shows that time-dependent calculations of Ng, Reames, and Tylka (2012) also exceed these limits because there has not yet been enough proton flow to establish equilibrium at the highest energies.

D. V. Reames and C. K. Ng

## 5. Spatial Dependence and Time Evolution

So far we have emphasized observations near 1 AU, but the evolution of the wave growth in space and time is quite complex. Figure 9 shows the typical evolution of low-energy protons in radius and time in a large SEP event.

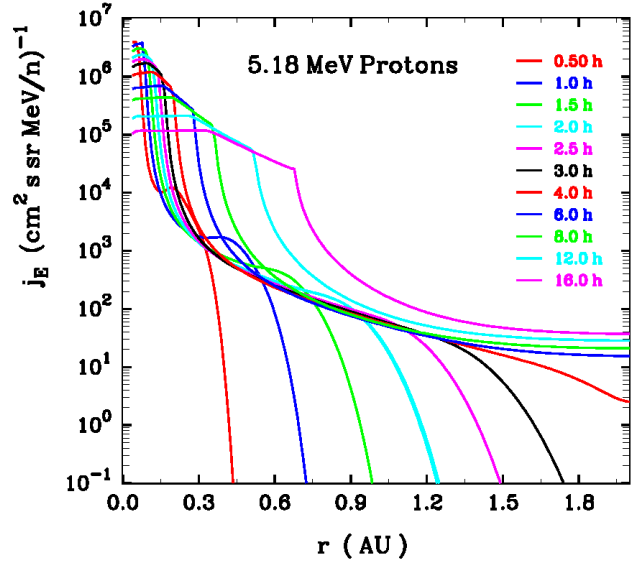

**Figure 9.** Intensities of 5.18 MeV protons are shown as a function of radius and time. Soon after arrival at a given radius, intensities rise to the streaming limit and remain there until the approach of the shock when they rise to the higher level. The cluster of curves along the streaming limit rises sharply from ~$10^2$ (cm$^2$ s sr MeV)$^{-1}$ at 1 AU to over $10^6$ (cm$^2$ s sr MeV)$^{-1}$ near the Sun.

The expected variations in wave intensities are illustrated by the scattering mean free path, $\lambda$ vs. $P$ at 0.35 AU at various times as shown in Figure 10. Clearly, models treating diffusion coefficients as constant are highly approximate.

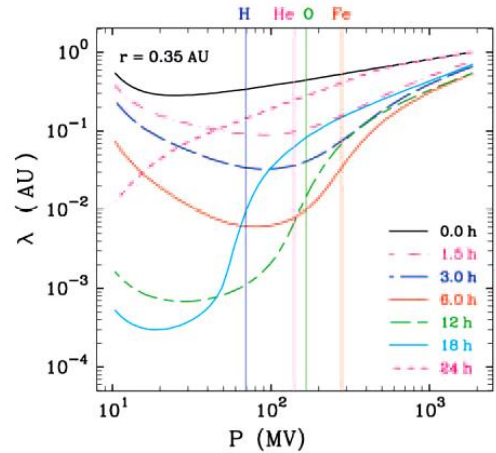

**Figure 10.** The scattering mean free path, $\lambda$ is shown vs. rigidity $P$ at various times at 0.35 AU. Vertical lines show the relative rigidities of H, He, O and Fe at constant velocity. Note that the scattering can vary by orders of magnitude. However, the variations are somewhat more modest farther from the Sun (see Ng, Reames, and Tylka 2003).

For SEP events near central meridian on the Sun, the duration of the streaming-limited plateau (identified in Figure 11) varies inversely as the shock speed. The more particles trapped by the streaming limit, the higher the shock peak will be. However, *it is impossible to predict the intensities at the shock peak from measurements of the earlier intensities on the streaming-limited plateau.*

The Streaming Limit

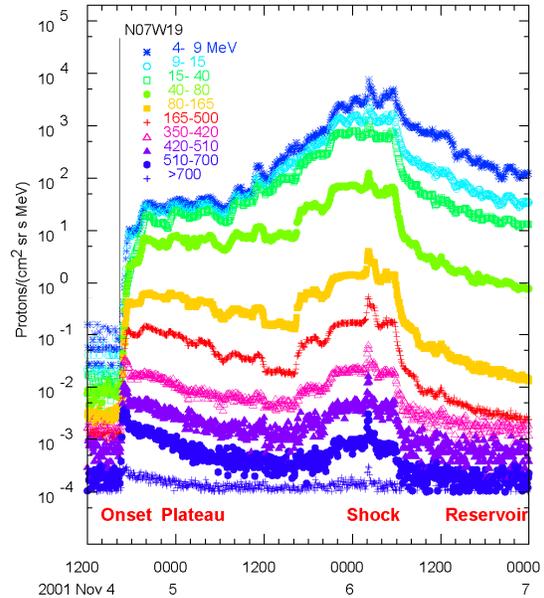

**Figure 11**. GOES data for the SEP event of 4 November 2001 are shown for proton energies up to >700 MeV as indicated. Phases of the event are shown along the bottom of the figure (see Reames 2013). Particles trapped by the streaming limited plateau will pile up near the shock. The resulting shock peak can be arbitrarily high.

## 6. Energy Spectral Knees

Related to the wave-particle physics that produces the streaming limit is the formation of energy spectral knees, sudden steepening of the particle energy spectra at high energies. These occur when the particle rigidity is reached where the resonant waves are becoming inadequate to contain the particles near the shock. For example, Lee (2005) places this at the balance point between scattering and focusing by the diverging magnetic field.

The evolution of proton spectra in the acceleration model of Ng and Reames (2008) is shown in the left panel of Figure 12. The right panel shows the shock-frame spatial distribution at 12.3 MeV finding a limit at ~0.1 solar radii.

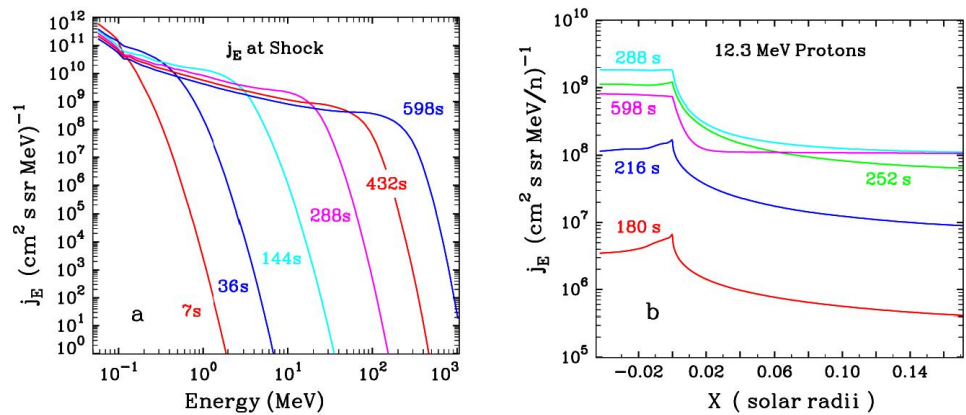

**Figure 12**. The left panel shows the time evolution of the proton spectrum at a shock. The right panel shows the radial distribution at 12.3 MeV attaining a streaming limit within 0.1 radii.

With increasing time the shock will begin to weaken and the spectrum will shrink back, leaving a break at high energies similar to those shown in Figure 10. The knee occurs at the highest energy attained before the shock weakens. The theory of proton energy spectral knees has been also discussed by Ellison and Ramaty (1985), Lee (2005), and Sandroos and Vainio (2009). Scaling of the knee to other ions has been studied by Tylka and Lee (2006), Li *et al.* (2009), and Battarbee, Laitinen, and Vainio (2011)

The radiation hazard to astronauts becomes difficult to shield, even with ~10 g cm$^{-2}$ of Al, for protons with energies above about 150 MeV. Proton spectra with knee energies above a few hundred MeV are extremely dangerous. However, above ~500 MeV, intensities are usually too low to be a threat.

## 7. Summary

1) The streaming limit results from textbook plasma physics. Intense streaming protons cause growth of Alfvén waves that scatter the protons and reduce the streaming, establishing a wave-particle equilibrium.

2) An equilibrium is established between waves, $k$ and particles with resonant values of $\mu P$. This allows coupling where high-energy particles at small $\mu$ limit lower-energy particles with larger $\mu$. This can produce energy spectra that are flattened at low energy.

3) The equilibrium is established, or not, on each magnetic flux tube and may vary independently in solar latitude and longitude depending upon the spatial properties of the magnetically-connected shock source.

4) A minimum fluence of protons is required to generate enough waves to establish the equilibrium. In a fast rising SEP event the proton intensity may overshoot the streaming limit briefly until the equilibrium is established.

5) The streaming limit is a transport phenomenon. It does *not* depend upon the nature of the particle source. For example, quasi-parallel or quasi-perpendicular shock waves producing the same $\delta f/\delta \mu$ will produce the same wave growth and will attain the same streaming limit.

6) Energy spectral knees are a high-energy limit of the wave-particle balance at the shock. A better knowledge of the parameters that determine the proton spectral knee is important for predicting radiation hazards.

The Streaming Limit

7) The wave-particle physics is dominated by protons; heavier ions respond to the resonant waves.

This paper was presented at the workshop on Extreme Space Weather Events in Boulder, Co, June 9-11, 2014.